# Proofs for the New Definitions in Financial Markets

Atilla Aras[a]

[a] Gazi University, Graduate School of Natural and Applied Sciences

Department of Mathematics

Ankara, Türkiye

E-mail: aaras1974@gmail.com

ORCID ID number: 0000-0002-7727-9797

**Funding**: No funds, grants, or other support was received.

**Proofs for the New Definitions in Financial Markets**

**Abstract**

The aim of this study is to present proofs for new theorems. Basic thoughts of new definitions emerge from the decision-making under uncertainty in economics and finance. Shape of the certain utility curve is central to standard definitions in determining risk attitudes of investors. Shape alone determines risk behavior of investors in standard theory. Although the terms "risk-averse," "risk-loving," and "risk-neutral" are equivalent to "strict concavity," "strict convexity," and "linearity," respectively, in standard theory, strict concavity or strict convexity, or linearity are valid for certain new definitions. The connection between the curvature of utility curve and risk attitude is broken for the new definitions. For instance, convex utility curve may show risk-averse behavior under new definitions. Additionally, this paper has proved that new definitions are richer than standard ones when shape is considered. Hence, it can be stated that new definitions are broader than standard definitions from the viewpoint of shape. With all of these, it has been demonstrated that the theorems and proofs in this study extend the standard utility theory in an important way.

# Proofs for the New Definitions in Financial Markets

## 1. Background

Aras (2022, 2024) has recently introduced the new definitions to determine the risk attitudes of investors in financial markets. Aras has provided empirical evidence to support the validity of these definitions. The new definitions differ from those in the standard theory on the assumptions concerning the certain utility curves of the investors. There is no connection between the shape of the utility curve and risk attitude under the new definitions. The aim of this study is to determine the shapes of the certain utility curves of investors for the new definitions. The theorems detailed herein were constructed to achieve this aim. Hence, it can be concluded that there is a research gap in the literature about the shapes of the certain utility curves for the new definitions of various risk behaviors.

Certain utility curves describing investors are strictly concave, strictly convex, and linear, for the risk-averse, risk-loving, and risk-neutral investors, respectively, in standard theory. Now, we can know that some new definitions in Aras's work may satisfy strict concavity or strict convexity, or linearity, but not in the same way as standard theory. The new definitions possess the required shapes of the standard ones, and they have additional shapes as well. For instance, the shape of the certain utility curve of a risk-averse investor can be strictly convex under the new definitions. Although the shape of the certain utility curve of investors is a must in standard theory, some new definitions by Aras (2022, 2024) satisfy the aforementioned shapes one by one. Hence it can be concluded that new definitions have broader content than standard definitions when shape of the certain utility curve is taken into account. This provides scientists broader framework when working on investor behaviors.

The motivation of the study is that all risk attitudes of investors in financial markets (i.e., not enough (insufficient) risk-loving) cannot be included to the existing problems in most cases by standard theory. For instance, some risk attitudes of investors cannot be determined in the problem of equity premium puzzle by using standard theory. Hence, constructing the shapes of the certain utility curves in the new definitions is a first step to determine all risk attitudes of investors in financial markets for the problems in theory.

The study's contribution to the existing literature is determining the shapes of the certain utility curves of new definitions for financial markets which represent the risk attitudes in standard classical utility theory. Because I prove mathematically that new definitions are free from curvatures of utility curves, theorems and proofs in this study extend the classical utility theory.

## 2. Literature Review

New definitions of Aras (2022, 2024) are based on decision-making under risk. Decision-making under risk depends on the shape of certain utility curve in standard utility theory.

There is a vast amount of literature about decision-making under risk. Many scientists have been working on this topic and new theories are still being formulated. The Expected Utility Theory (EUT) is the oldest of these. It was first proposed by Bernoulli (1738) to address the problem of how much a rational person pays when gambling. The answer is the expected monetary value of the gamble. Bernoulli gave a counterexample, known as the St. Petersburg paradox. It is as follows: although the expected payoff of a lottery approaches infinity, the lottery itself is worth a very small amount to the game's players.

Von Neumann and Morgenstern (1944) addressed the same problem and developed the EUT theory. The theory depends on preferences and axioms. Savage (1954) proposed another theory, the Subjective Expected Utility Theory (SEUT), about decision-making under risk. It builds on the basic ideas of Ramsey (1931), De Finetti (1937), and Von Neumann and Morgenstern (1944). The main difference between the EUT and SEUT is how the theories treat probability. In the EUT, probability is based on objective information, whereas in the SEUT, it is perceived subjectively by decision-makers.

Economists have tried to separate the decision-making under risk theories into different categories (Suhonen, 2007). The separation of normative and descriptive theories is one of these. Normative theories are interested in how people should behave in risky environments. Descriptive theories, by contrast, are focused on how people make decisions in real environments. The Weighted Utility Theory (Chew and MacCrimmon, 1979) and Generalized Expected Utility Theory (Machina, 1982) are examples of the descriptive theory. Starmer (2000) explains the common features of these theories as follows:

1. Preferences are defined over prospects.
2. Their functions satisfy continuity and ordering.
3. The principle of monotonicity is followed.

Regret theory is another example of a descriptive theory. It was proposed simultaneously by Bell (1982), Fishburn (1982), and Loomes and Sugden (1982, 1987). The theory states that the fear of regret may preclude people from taking action, or it may lead a person take action. Investors in financial markets are affected so that they become unnecessarily risk-adverse or risk-loving.

Kahneman and Tversky (1992, 2013) formulated the Prospect Theory, which is another example of descriptive theory. Their approach to decision-making under risk was to refer to traditional behavioral sciences. Their experimental findings (1981, 1986) were the backbones for their theory. They posited that people value losses and gains differently and make decisions based on perceived gains, rather than perceived losses.

Hey and Orme (1994), Harless and Camerer (1994), and Loomes and Sugden (1995) developed the models in the Theory of Stochastic Preference. This is another alternative descriptive theory. Models of this theory have common features; for instance, they share deterministic core theories of preference.

Studies on decision-making under uncertainty continued in the last five years. Walters et al. (2023) showed that investor behavior is sensitive to epistemic uncertainty and aleatory uncertainty. They stated that these two different kinds of uncertainty shape investor behavior. Tong et al. (2023) studied on how financial volatility react to economic uncertainty. They found that simultaneous effects of uncertainty on volatility are positive. Campello and Kankanhalli (2024) investigated how different uncertainties affect corporate decision-making. Additionally, Bai et al. (2025) studied the use of reinforcement learning to aid sequential decision-making under uncertainty. Finally, Buehler et al. (2025) suggested a method based on subsampling to express model uncertainty.

EUT has been heavily criticized by scientists since the early 1950s. Scientists criticize it because decision-makers systematically violate the rationality axioms. Allais' paradox (1953) is one criticism of the EUT. Other problems and paradoxes against EUT are preference reversal, as explained by Lichtenstein and Slovic (1971); Ellsberg's paradox, by Ellsberg (1961); and other experiments that indicate that decision-makers violate rationality axioms.

Many alternative models to EUT have been developed since the mid to late twentieth century, and many scientists expect this to continue. Because none of the alternative theories presented to date have been able to express all paradoxes and to overcome the aforementioned problems, the EUT cannot be rejected and replaced. At the same time, it would be illogical to reject all alternative theories. Hence, scientists propose that there should be one core theory (i.e., EUT) with alternative theories (Suhonen, 2007).

Scientists also continued to work on the shapes of certain utility curves and their connection to risk over the last five years. Aras (2022, 2024) introduced new definitions for financial markets in his studies. He also coined the term "sufficiency factor of the model". This new term helped to solve the equity premium puzzle. Bleichrodt et al. (2023) revealed that Hurwicz Expected Utility is compatible with empirical data and it can be tested. Goeree and Garcia-Pola (2023) made use of non-parametric tests to show that there is a deviation from concave utility models in the real world. Liu and Shen (2024) investigated PSAHARA and SAHARA and demonstrated that individuals are less risk-averse when absolute risk aversion is not monotonic. Phelps (2024) reviewed the shape of CRRA and CARA utility functions. He also stated that generalized logistic utility involves both concave and convex parts. In contrast, Liang et al. (2024) showed that piecewise hyperbolic absolute risk aversion utilities possess non-concave parts and change optimal allocation of portfolios by enlarging risk-seeking. Additionally, kinks of these utilities lead lower risk-taking. Levy (2024) demonstrated that utility curves should be almost logarithmic to possess consistent decision making. Finally, meta-analysis of Elmirejad et al. (2025) revealed that relative risk aversion in economics field is approximately 1. In contrast, the analysis showed that estimations in the finance area are in the range of 2-7. The study also expressed that estimates vary according to the type of context.

## 3. Materials & Methods

### 3.1 Definitions

Aras (2022, 2024) has introduced some definitions concerning the financial markets. These differ on the assumption of the certain utility curves of the investors. This paper proved that his new definitions are free from the curvatures of the utility curves. For example, strict concave utility curve may show the risk-neutral behavior of investors under new definitions. This leads a great flexibility for scientists working on risk.

He (2022) also coined the term “the sufficiency factor of the model”. His new definitions involve this term. Sufficiency factor of the model is a coefficient for uncertain utilities to adjust uncertainty to make possible to compare certain and uncertain utility and leads a solution for the equity premium puzzle.

### 3.2 Theorems

The first three theorems are standard in theory. However, the last five theorems and two propositions were constructed by me (Theorems 4, 5, 6, 7, and 8).

**Theorem 1.** Suppose that $v(w)$ is an increasing curve.

The agent is risk-averse in standard form ⟺ $v(w)$ is strictly concave.

**Theorem 2.** Suppose that $v(w)$ is an increasing curve.

The agent is risk-loving in standard form ⟺ $v(w)$ is strictly convex.

**Theorem 3.** Suppose that $v(w)$ is an increasing curve.

The agent is risk-neutral in standard form ⟺ $v(w)$ is linear.

**Proposition 1.** Suppose that $v(w)$ is an increasing strict convex utility curve for the agent on a given interval and variable $w_{t+1}$ denotes the wealth at current time. Let $w_{t+2}$ - $w_{t+1}$ be less than ∞. Inequality $Ev(w_{t+2}) > v(w_{t+1})$ holds true for the agent.

*Proof.* Let $v(w)$ be an increasing strict convex utility curve for the agent. Because the utility curve is an increasing strict convex curve, inequalities $w_{t+2} > w_{t+1}$ and $v(w_{t+2}) < Ev(w_{t+2})$ hold true. Hence, we possess $v(w_{t+2}) - v(w_{t+1})/\infty < v(w_{t+2}) - v(w_{t+1})/w_{t+2} - w_{t+1} < Ev(w_{t+2}) - v(w_{t+1})/w_{t+2} - w_{t+1}$. Thus, $Ev(w_{t+2}) - v(w_{t+1})/w_{t+2} - w_{t+1} > 0$. Because $w_{t+2} - w_{t+1} > 0$, $Ev(w_{t+2}) - v(w_{t+1}) > 0$. Therefore, $Ev(w_{t+2}) > v(w_{t+1})$ holds true for the agent.

Q.E.D.

**Proposition 2.** Suppose that $v(w)$ is an increasing strict concave utility curve for the agent on a given interval and variable $w_{t+1}$ denotes the wealth at current time. Let $w_{t+2}$ - $w_{t+1}$ be less than ∞. Inequalities $Ev(w_{t+2}) > v(w_{t+1}), Ev(w_{t+2}) < v(w_{t+1})$, or $Ev(w_{t+2}) = v(w_{t+1})$ hold true for the agent.

*Proof.* Let $v(w)$ be an increasing strict concave utility curve for the agent. Because the utility curve is an increasing strict concave curve, inequalities $w_{t+2} > w_{t+1}$ and $v(w_{t+2}) > Ev(w_{t+2})$ hold true. Hence, we possess the following inequality. $Ev(w_{t+2}) - v(w_{t+1})/\infty < Ev(w_{t+2}) - v(w_{t+1})/w_{t+2} - w_{t+1} < v(w_{t+2}) - v(w_{t+1})/w_{t+2} - w_{t+1}$. Thus, inequalities $Ev(w_{t+2}) > v(w_{t+1}), Ev(w_{t+2}) < v(w_{t+1})$, or $Ev(w_{t+2}) = v(w_{t+1})$ hold true for the agent.

Q.E.D.

**Theorem 4.** Suppose that $v(w)$ is an increasing utility curve on a given interval and the agent's subjective time discount factor ($\beta$) has an acceptable value. Variable $w_{t+1}$ denotes the wealth at current time. The agent is allocating extra negative utility for the uncertain wealth value (i.e., $\Omega_{t+1} < 1$, that is, sufficiency factor of the model is less than 1)

The following holds true on the given interval for nontrivial lotteries with some $\beta\Omega_{t+1} < 1$ values.

The agent is risk-averse or risk-loving, or risk-neutral in standard forms ⇔ $\beta\Omega_{t+1}Ev(w_{t+2}) < v(w_{t+1})$

*Proof.* (⟹)

Suppose that $u(w)$ is an increasing utility curve with some $\beta\Omega_{t+1} < 1$ values.

*Case 1.* Suppose that $Ev(w_{t+1}) < v[E(\mathrm{w})] = v(w_{t+1})$ on the given interval with some $\beta\Omega_{t+1} < 1$ values. Inequality $Ev(w_{t+1}) < v[E(\mathrm{w})] = v(w_{t+1})$ is equal to strict concavity. Thus, choose an arbitrary probability value p with 0 < p < 1 for the expected value. We have three alternatives for the increasing utility curve by proposition 2.

*Alternative 1.* We have $Ev(w_{t+2}) > v(w_{t+1})$ by proposition 2. Because there exist some $\beta\Omega_{t+1} < 1$ values, such that $\beta\Omega_{t+1}Ev(w_{t+2}) < Ev(w_{t+1}) < v[E(w)] = v(w_{t+1}) <$ $Ev(w_{t+2})$ holds true, $Ev(w_{t+1}) < v[E(\text{w})] = v(w_{t+1})$ implies $\beta\Omega_{t+1}Ev(w_{t+2}) < v(w_{t+1})$ with some $\beta\Omega_{t+1} < 1$ values.

*Alternative 2.* We have $Ev(w_{t+2}) < v(w_{t+1})$ by proposition 2. Because there exist some $\beta\Omega_{t+1} < 1$ values, such that $\beta\Omega_{t+1}Ev(w_{t+2}) < Ev(w_{t+1}) < Ev(w_{t+2}) < v[E(w)] = v(w_{t+1})$ holds true, $Ev(w_{t+1}) < v[E(\text{w})] = v(w_{t+1})$ implies $\beta\Omega_{t+1}Ev(w_{t+2}) < v(w_{t+1})$ with some $\beta\Omega_{t+1} < 1$ values.

*Alternative 3.* We have $Ev(w_{t+2}) = v(w_{t+1})$ by proposition 2. Because there exist some $\beta\Omega_{t+1} < 1$ values, such that $\beta\Omega_{t+1}Ev(w_{t+2}) < Ev(w_{t+1}) < Ev(w_{t+2}) = v[E(w)] = v(w_{t+1})$ holds true, $Ev(w_{t+1}) < v[E(\text{w})] = v(w_{t+1})$ implies $\beta\Omega_{t+1}Ev(w_{t+2}) < v(w_{t+1})$ with some $\beta\Omega_{t+1} < 1$ values.

*Case 2.* Suppose that $v[E(\text{w})] = v(w_{t+1}) < Ev(w_{t+1})$ on the given interval with some $\beta\Omega_{t+1} < 1$ values. Inequality $v[E(\text{w})] = v(w_{t+1}) < Ev(w_{t+1})$ is equal to strict convexity. Thus, choose an arbitrary probability value p with 0 < p < 1 for the expected value. We have $v(w_{t+1}) < Ev(w_{t+2})$ for the increasing utility curve by proposition 1. Because there exist some $\beta\Omega_{t+1} < 1$ values, such that $\beta\Omega_{t+1}Ev(w_{t+2}) < v[E(\text{w})] = v(w_{t+1}) < Ev(w_{t+1}) < Ev(w_{t+2})$ holds true, $v[E(\text{w})] = v(w_{t+1}) < Ev(w_{t+1})$ implies $\beta\Omega_{t+1}Ev(w_{t+2}) < v(w_{t+1})$ with some $\beta\Omega_{t+1} < 1$ values.

*Case 3.* Suppose that $v[E(\text{w})] = v(w_{t+1}) = Ev(w_{t+1})$ on the given interval with some $\beta\Omega_{t+1} < 1$ values. Inequality $v[E(\text{w})] = v(w_{t+1}) = Ev(w_{t+1})$ is equal to linearity. Thus, choose an arbitrary probability value p with 0 ≤ p ≤ 1 for the expected value. There are three conditions to examine.

*Condition 1.* Choose p = 1 for the expected value. Moreover, $v[E(\text{w})] = v(w_{t+1}) = Ev(w_{t+1})$ holds true. Because there exist all $\beta\Omega_{t+1} < 1$ values, such that $\beta\Omega_{t+1}Ev(w_{t+2}) < v(w_{t+1}) = v[E(\text{w})] = Ev(w_{t+1}) = Ev(w_{t+2}) = v(w_{t+2})$ holds true, $v[E(\mathrm{w})] = v(w_{t+1}) = Ev(w_{t+1})$ implies $\beta\Omega_{t+1}Ev(w_{t+2}) < v(w_{t+1})$ with some $\beta\Omega_{t+1} < 1$ values.

*Condition 2.* Choose p with $0 < \mathrm{p} < 1$ for the expected value. Then, we have $v[E(\mathrm{w})] = v(w_{t+1}) = Ev(w_{t+1}) < Ev(w_{t+2})$ for the increasing utility curve. Because there exist some $\beta\Omega_{t+1} < 1$ values, such that $\beta\Omega_{t+1}Ev(w_{t+2}) < v[E(\mathrm{w})] = v(w_{t+1}) = Ev(w_{t+1}) < Ev(w_{t+2})$ holds true, $v[E(\mathrm{w})] = v(w_{t+1}) = Ev(w_{t+1})$ implies $\beta\Omega_{t+1}Ev(w_{t+2}) < v(w_{t+1})$ with some $\beta\Omega_{t+1} < 1$ values.

*Condition 3.* Choose $\mathrm{p} = 0$ for the expected value. Then, $v[E(\mathrm{w})] = v(w_{t+1}) = Ev(w_{t+1}) = v(w_t) = Ev(w_t) < Ev(w_{t+2})$ for the increasing utility curve. Because there exist some $\beta\Omega_{t+1} < 1$ values, such that $\beta\Omega_{t+1}Ev(w_{t+2}) < v[E(\mathrm{w})] = v(w_{t+1}) = Ev(w_{t+1}) = v(w_t) = Ev(w_t) < Ev(w_{t+2})$ holds true, $v[E(\mathrm{w})] = v(w_{t+1}) = Ev(w_{t+1})$ implies $\beta\Omega_{t+1}Ev(w_{t+2}) < v(w_{t+1})$ with some $\beta\Omega_{t+1} < 1$ values.

Since we know $Ev(w_{t+1}) < v[E(\mathrm{w})] = v(w_{t+1})$, $v[E(\mathrm{w})] = v(w_{t+1}) < Ev(w_{t+1})$, or $v[E(\mathrm{w})] = v(w_{t+1}) = Ev(w_{t+1})$ with some $\beta\Omega_{t+1} < 1$ values, these cases cover all the possibilities, so we can conclude that $\beta\Omega_{t+1}Ev(w_{t+2}) < v(w_{t+1})$ holds true with some $\beta\Omega_{t+1} < 1$ values.

($\Leftarrow$)

Suppose that $\beta\Omega_{t+1}Ev(w_{t+2}) < v(w_{t+1})$ holds true on the given interval with some $\beta\Omega_{t+1} < 1$ values and $v(w)$ is an increasing curve. Now again assume that the agent is not risk-averse and is not risk-neutral in standard forms. This assumption means that we suppose $v(w_{t+1}) = v[E(\mathrm{w})] \leq Ev(w_{t+1})$ and $v(w_{t+1}) = v[E(\mathrm{w})] \neq Ev(w_{t+1})$. Hence $\beta\Omega_{t+1}Ev(w_{t+2}) < v(w_{t+1}) = v[E(\mathrm{w})] < Ev(w_{t+1})$ holds true with some $\beta\Omega_{t+1} < 1$ values. We can infer from the last inequality that $v(w_{t+1}) = v[E(\mathrm{w})] < Ev(w_{t+1})$ holds true. Then the agent is said to be risk-loving in standard form. Hence, we can conclude that the agent is risk-averse or risk-loving, or risk-neutral in standard forms with some $\beta\Omega_{t+1} < 1$ values.

Q.E.D.

**Theorem 5.** Suppose that $v(w)$ is an increasing utility curve on a given interval. The agent's subjective time discount factor ($\beta$) has an acceptable value. Variable $w_{t+1}$ denotes the wealth at current time. The agent is allocating extra positive utility for the uncertain wealth value (i.e., $\Omega_{t+1} > 1$).

The following holds true on the given interval for nontrivial lotteries with some $\beta\Omega_{t+1} \geq 1$ values.

The agent is risk-averse or risk-loving, or risk-neutral in standard forms $\Leftrightarrow v(w_{t+1}) < \beta\Omega_{t+1}Ev(w_{t+2})$.

*Proof.* ($\Longrightarrow$)

That the agent is allocating extra positive utility for the uncertain wealth value is equivalent to $\beta\Omega_{t+1} \geq 1$ values. Suppose that $v(w)$ is an increasing utility curve with some $\beta\Omega_{t+1} \geq 1$ values.

*Case 1.* Suppose that $Ev(w_{t+1}) < v[E(\mathrm{w})] = v(w_{t+1})$ on the given interval with some $\beta\Omega_{t+1} \geq 1$ values. Then there exist some $\beta\Omega_{t+1} \geq 1$ values, such that there is one condition to examine.

*Condition 1.* Choose $0 < \mathrm{p} < 1$. Then, $Ev(w_{t+1}) < v[E(\mathrm{w})] = v(w_{t+1})$, which is equal to strict concavity, implies three alternatives by proposition 2.

*Alternative 1.* We have $Ev(w_{t+2}) > v(w_{t+1})$ by proposition 2. Because there exist all $\beta\Omega_{t+1} \geq 1$ values, $Ev(w_{t+1}) < v[E(\mathrm{w})] = v(w_{t+1}) < Ev(w_{t+2}) \leq \beta\Omega_{t+1}Ev(w_{t+2})$ holds true. Thus $Ev(w_{t+1}) < v[E(\mathrm{w})] = v(w_{t+1})$ implies $v(w_{t+1}) < \beta\Omega_{t+1}Ev(w_{t+2})$ with some $\beta\Omega_{t+1} \geq 1$ values.

*Alternative 2.* We have $Ev(w_{t+2}) < v(w_{t+1})$ by proposition 2. Because there exist some $\beta\Omega_{t+1} \geq 1$ values, $Ev(w_{t+1}) < Ev(w_{t+2}) < v[E(\mathrm{w})] = v(w_{t+1}) \leq \beta\Omega_{t+1}Ev(w_{t+2})$ holds true. Moreover, former inequality implies $Ev(w_{t+1}) < Ev(w_{t+2}) < v[E(\mathrm{w})] = v(w_{t+1}) <$ $\beta\Omega_{t+1}Ev(w_{t+2})$. Thus, $Ev(w_{t+1}) < v[E(\mathrm{w})] = v(w_{t+1})$ implies $v(w_{t+1}) < \beta\Omega_{t+1}Ev(w_{t+2})$ with some $\beta\Omega_{t+1} \geq 1$ values.

*Alternative 3.* We have $Ev(w_{t+2}) = v(w_{t+1})$ by proposition 2. Because there exist all $\beta\Omega_{t+1} \geq 1$ values, $Ev(w_{t+1}) < Ev(w_{t+2}) = v(w_{t+1}) = v[E(\mathrm{w})] \leq \beta\Omega_t Ev(w_{t+2})$ holds true. Moreover, former inequality implies $Ev(w_{t+1}) < Ev(w_{t+2}) = v(w_{t+1}) = v[E(\mathrm{w})] < \beta\Omega_t Ev(w_{t+2})$ Thus, $Ev(w_{t+1}) < v[E(\mathrm{w})] = v(w_{t+1})$ implies $v(w_{t+1}) < \beta\Omega_t Ev(w_{t+2})$ with some $\beta\Omega_{t+1} \geq 1$ values.

*Case 2.* Suppose that $v[E(\mathrm{w})] = v(w_{t+1}) < Ev(w_{t+1})$ on the given interval with some $\beta\Omega_{t+1} \geq 1$ values. Inequality $v[E(\mathrm{w})] = v(w_{t+1}) < Ev(w_{t+1})$ is equal to strict convexity. Then, $v(w_{t+1}) = v[E(\mathrm{w})] < Ev(w_{t+2})$ holds true by proposition 1. We can choose all $\beta\Omega_{t+1} \geq 1$ values, such that $v(w_{t+1}) = v[E(\mathrm{w})] < Ev(w_{t+1}) < Ev(w_{t+2}) \leq \beta\Omega_{t+1}Ev(w_{t+2})$ holds true. Thus, $v[E(\mathrm{w})] = v(w_{t+1}) < Ev(w_{t+1})$ implies $v(w_{t+1}) < \beta\Omega_{t+1}Ev(w_{t+2})$ with some $\beta\Omega_{t+1} \geq 1$ values.

*Case 3.* Suppose that $v(w_{t+1}) = v[E(\mathrm{w})] = Ev(w_{t+1})$ on the given interval with some $\beta\Omega_{t+1} \geq 1$ values. Inequality $v[E(\mathrm{w})] = v(w_{t+1}) = Ev(w_{t+1})$ is equal to linearity. We can choose all $\beta\Omega_{t+1} \geq 1$ values, such that $v(w_{t+1}) = v[E(\mathrm{w})] = Ev(w_{t+1}) < Ev(w_{t+2}) \leq \beta\Omega_{t+1}Ev(w_{t+2})$ holds true. Thus, $v(w_{t+1}) = v[E(w_{t+1})] = Ev(w_{t+1})$ implies $v(w_{t+1}) < \beta\Omega_{t+1}Ev(w_{t+2})$ with some $\beta\Omega_{t+1} \geq 1$ values.

Since we know $Ev(w_{t+1}) < v(w_{t+1}) = v[E(\mathrm{w})]$, $v(w_{t+1}) = v[E(\mathrm{w})] < Ev(w_{t+1})$, or $v(w_{t+1}) = v[E(\mathrm{w})] = Ev(w_{t+1})$ with some $\beta\Omega_{t+1} \geq 1$ values, these cases cover all the possibilities, so we can conclude that $v(w_{t+1}) < \beta\Omega_{t+1}Ev(w_{t+2})$ holds true with some $\beta\Omega_{t+1} \geq 1$ values.

($\Leftarrow$)

Suppose that $v(w_{t+1}) < \beta\Omega_{t+1}Ev(w_{t+2})$ holds true on the given interval with some $\beta\Omega_{t+1} \geq 1$ values and $v(w)$ is an increasing curve. Now again assume that the agent is not risk-loving and is not risk-neutral in standard forms. This assumption means that we suppose $v(w_{t+1})$ = $v[E(\mathrm{w})] \geq Ev(w_{t+1})$ and $v(w_{t+1})$ = $v[E(\mathrm{w})] \neq Ev(w_{t+2})$. Hence $Ev(w_{t+1}) < v(w_{t+1})$ = $v[E(\mathrm{w})]$ < $\beta\Omega_{t+1}Ev(w_{t+2})$ holds true with all $\beta\Omega_{t+1}$ ≥1 values. We can infer from the last inequality that $Ev(w_{t+1}) < v(w_{t+1})$ = $v[E(\mathrm{w})]$ holds true some $\beta\Omega_{t+1}$ ≥1 values. Then the agent is said to be risk-averse in standard form. Hence, we can conclude that the agent is risk-averse or risk-loving, or risk-neutral in standard forms with some $\beta\Omega_{t+1} \geq 1$ values.

Q.E.D.

**Theorem 6.** Suppose that $v(w)$ is an increasing utility curve on a given interval. The agent's subjective time discount factor ($\beta$) has an acceptable value. Variable $w_{t+1}$ denotes the wealth at current time. The agent is allocating extra positive utility for the uncertain wealth value (i.e., $\Omega_{t+1} > 1$).

The following holds true on the given interval for nontrivial lotteries with some $\beta\Omega_{t+1} \leq 1$ values.

The agent is risk-loving or risk-neutral in standard forms ⟺ $v(w_{t+1}) < \beta\Omega_{t+1}Ev(w_{t+2})$.

*Proof.* (⟹)

That the agent is allocating extra positive utility for the uncertain wealth value is equivalent to $\beta\Omega_{t+1} \leq 1$. Suppose that $v(w)$ is an increasing utility curve some $\beta\Omega_{t+1} \leq 1$ values.

*Case 1.* Suppose that $v(w_{t+1}) = v[E(\mathrm{w})] < Ev(w_{t+1})$ on the given interval. Inequality $v(w_{t+1}) = v[E(\mathrm{w})] < Ev(w_{t+1})$ is equal to strict convexity. There exist some $\beta\Omega_{t+1} \leq 1$ values, such that $v(w_{t+1}) = v[E(\mathrm{w})] < Ev(w_{t+1}) < \beta\Omega_{t+1}Ev(w_{t+2}) \leq Ev(w_{t+2})$ holds true. Thus, $v(w_{t+1}) = v[E(\mathrm{w})] < Ev(w_{t+1})$ implies $v(w_{t+1}) < \beta\Omega_{t+1}Ev(w_{t+2})$ with some $\beta\Omega_{t+1} \leq 1$ values.

*Case 2.* Suppose that $v(w_{t+1}) = v[E(\text{w})] = Ev(w_{t+1})$ on the given interval. Inequality $v(w_{t+1}) = v[E(\text{w})] = Ev(w_{t+1})$ is equal to linearity. There exist some $\beta\Omega_{t+1} \leq 1$ values, such that $v(w_{t+1}) = v[E(\text{w})] = Ev(w_{t+1}) < \beta\Omega_{t+1}\, Ev(w_{t+2}) \leq Ev(w_{t+2})$ holds true. Thus, $v(w_{t+1}) = v[E(\text{w})] = Ev(w_{t+1})$ implies $v(w_{t+1}) < \beta\Omega_{t+1}Ev(w_{t+2})$ with some $\beta\Omega_{t+1} \leq 1$ values.

Since we know $v(w_{t+1}) = v[E(\text{w})] < Ev(w_{t+1})$ or $v(w_{t+1}) = v[E(\text{w})] = Ev(w_{t+1})$ with some $\beta\Omega_{t+1} \leq 1$ values, these cases cover all the possibilities, so we can conclude that $v(w_{t+1}) < \beta\Omega_{t+1}Ev(w_{t+2})$ with some $\beta\Omega_{t+1} \leq 1$ values.

($\Longleftarrow$)

Suppose that $v(w_{t+1}) < \beta\Omega_{t+1}Ev(w_{t+2})$ holds true with some $\beta\Omega_{t+1} \leq 1$ values and $v(w)$ is an increasing utility curve. Now assume that the agent is not risk-loving. This statement means that $v(w_{t+1}) = v[E(\text{w})] \geq Ev(w_{t+1})$. The last inequality implies that $v(w_{t+1}) = v[E(\text{w})] = Ev(w_{t+1}) \leq \beta\Omega_{t+1}Ev(w_{t+2}) \leq Ev(w_{t+2})$ with some $\beta\Omega_{t+1} \leq 1$ values. Hence the agent is said to be risk-neutral. We can conclude that the agent is risk-loving or risk-neutral in standard forms with some $\beta\Omega_{t+1} \leq 1$ values.

Q.E.D.

**Theorem 7.** Suppose that $v(w)$ is an increasing utility curve on a given interval. The agent's subjective time discount factor ($\beta$) has an acceptable value. Variable $w_{t+1}$ denotes the wealth at current time. The agent is allocating extra positive utility for the uncertain wealth value (i.e., $\Omega_{t+1} > 1$).

The following holds true on the given interval for nontrivial lotteries with some $\beta\Omega_{t+1} \leq 1$ values.

$v(w_{t+1}) < \beta\Omega_{t+1}Ev(w_{t+2})$ holds true $\Rightarrow$ The agent is risk-averse in standard form.

*Proof.*

That the agent is allocating extra positive utility for the uncertain wealth value is equivalent to $\beta\Omega_{t+1} \leq 1$. Suppose that $v(w_{t+1}) < \beta\Omega_{t+1}Ev(w_{t+2})$ holds true with an increasing utility curve and some $\beta\Omega_{t+1} \leq 1$ values. We can choose some $\beta\Omega_{t+1} \leq 1$ values, such that $Ev(w_{t+1}) < v(w_{t+1}) = v[E(\mathrm{w})] < \beta\Omega_{t+1}Ev(w_{t+2}) \leq Ev(w_{t+2})$ holds true. Hence, $Ev(w_{t+1}) < v(w_{t+1}) = v[E(\mathrm{w})]$ which is equal to strict concavity holds true.

Q.E.D.

**Theorem 8.** Suppose that $v(w)$ is an increasing curve on a given interval and the agent's subjective time discount factor ($\beta$) has an acceptable value. Variable $w_{t+1}$ denotes the wealth at current time. The agent is allocating extra positive or extra negative utility for the uncertain wealth value.

The following holds true on the given interval for nontrivial lotteries with some $\beta\Omega_{t+1} \leq 1$ values.

The agent is risk-averse or risk-loving, or risk-neutral in standard forms $\Leftrightarrow$ $\beta\Omega_{t+1}Ev(w_{t+2})$ $= v(w_{t+1})$.

*Proof.* ($\Longrightarrow$)

Some $\beta\Omega_{t+1} \leq 1$ values denote that the agent may allocate extra positive or extra negative utility for the uncertain wealth value. Suppose that $v(w)$ is an increasing utility curve with some $\beta\Omega_{t+1} \leq 1$ values.

*Case 1.* Suppose that $Ev(w_{t+1}) < v[E(\mathrm{w})] = v(w_{t+1})$ holds true on the given interval. Inequality $Ev(w_{t+1}) < v[E(\mathrm{w})] = v(w_{t+1})$ is equal to strict concavity. Because there exist some $\beta\Omega_{t+1} \leq 1$ values, such that $Ev(w_{t+1}) < \beta\Omega_{t+1}Ev(w_{t+2}) \leq Ev(w_{t+2}) = v[E(\mathrm{w})] = v(w_{t+1})$ holds true, smaller than or equality in this inequality implies equality. Hence, $Ev(w_{t+1}) < v[E(\mathrm{w})] = v(w_{t+1})$ implies $\beta\Omega_{t+1}Ev(w_{t+2}) = v(w_{t+1})$ with some $\beta\Omega_{t+1} \leq 1$ values.

*Case 2.* Suppose that $v[E(\mathrm{w})] = v(w_{t+1}) < Ev(w_{t+1})$ on the given interval. Inequality $v[E(\mathrm{w})] = v(w_{t+1}) < Ev(w_{t+1})$ is equal to strict convexity. Because there exist some $\beta\Omega_{t+1} \leq 1$ values, such that $\beta\Omega_{t+1}Ev(w_{t+2}) = v(w_{t+1}) = v[E(\mathrm{w})] \leq Ev(w_{t+1}) \leq Ev(w_{t+2})$ holds true, convexity in this inequality implies strict convexity. Hence, $v[E(\mathrm{w})] = v(w_{t+1}) < Ev(w_{t+1})$ implies $\beta\Omega_{t+1}Ev(w_{t+2}) = v(w_{t+1})$ with some $\beta\Omega_{t+1} \leq 1$ values.

*Case 3.* Suppose that $v[E(\mathrm{w})] = v(w_{t+1}) = Ev(w_{t+1})$ on the given interval. Inequality $v[E(\mathrm{w})] = v(w_{t+1}) = Ev(w_{t+1})$ is equal to linearity. Because there exist all $\beta\Omega_{t+1} \leq 1$ values, such that $\beta\Omega_{t+1}Ev(w_{t+2}) = v[E(\mathrm{w})] = v(w_{t+1}) = Ev(w_{t+1}) \leq Ev(w_{t+2})$ holds true, $v[E(\mathrm{w})] = v(w_{t+1}) = Ev(w_{t+1})$ implies $\beta\Omega_{t+1}Ev(w_{t+2}) = v(w_{t+1})$ with some $\beta\Omega_{t+1} \leq 1$ values.

Since we know $Ev(w_{t+1}) < v[E(\mathrm{w})] = v(w_{t+1})$ or $v[E(\mathrm{w})] = v(w_{t+1}) < Ev(w_{t+1})$, or $v[E(\mathrm{w})] = v(w_{t+1}) = Ev(w_{t+1})$ with some $\beta\Omega_{t+1} \leq 1$ values, these cases cover all the possibilities, so we can conclude that $v(w_{t+1}) = \beta\Omega_{t+1}Ev(w_{t+2})$ with some $\beta\Omega_{t+1} \leq 1$ values.

($\Leftarrow$)

Suppose that $\beta\Omega_{t+1}Ev(w_{t+2}) = v(w_{t+1})$ on the given interval and $v(w)$ is an increasing utility curve with some $\beta\Omega_{t+1} \leq 1$ values. Assume again that the agent is not risk-averse and is not risk-neutral in standard forms. This assumption means that $v[E(\mathrm{w})] = v(w_{t+1}) \leq Ev(w_{t+1})$ and $v[E(\mathrm{w})] = v(w_{t+1}) \neq Ev(w_{t+1})$ holds true. Hence, we can infer that $\beta\Omega_{t+1}Ev(w_{t+2}) = v[E(\mathrm{w})] = v(w_{t+1}) < Ev(w_{t+1}) < Ev(w_{t+2})$ is true with some $\beta\Omega_{t+1} \leq 1$ values. Then the agent is said to be risk-loving in standard form because $v[E(\mathrm{w})] = v(w_{t+1}) < Ev(w_{t+1})$ holds true. Hence, we can conclude that the agent is risk-averse or risk-loving, or risk neutral in standard forms with some $\beta\Omega_{t+1} \leq 1$ values.

Q.E.D.

## 4. Results and Discussion

Aras (2022, 2024) formulated new definitions for the financial markets. As Aras (2022) assumed concave certain utility curves for all investors, he also allowed different kinds of certain utility curves for investors (2024).

The shape of the certain utility curve is necessary for determining the risk attitudes of investors in standard theory. This is not the case for some new definitions. For instance, the shape of the certain utility curve of a risk-averse investor may be strictly convex under new definitions. Hence, we can conclude that new definitions are broader than standard definitions from the viewpoint of shape.

The new definitions are not equivalent to one another because each biconditional in the proofs has a different assumption (i.e., the agent is allocating extra utility differently for each biconditional). This situation is compatible with financial economics theory.

It is very difficult to compare certain and uncertain values at the same wealth value in one financial market in the real world as it is done in standard theory. Hence, it is much easier to determine the risk attitudes of investors in financial markets using these new definitions because there is no need to compare certain and uncertain utility curves at the same wealth value with them. New definitions are independent from the curvatures of utility curves. This property of new definitions both provides greater flexibility for scientists when modelling risk attitudes and extends the classical utility theory.

## 5. Conclusion

This paper has proved that certain new definitions for utility curves used by some types of investors may satisfy strict concavity or strict convexity, or linearity, which is not being the same way as standard theory. New definitions possess the required shapes of standard definitions and contain more shapes than standard ones when shape is considered. Hence, it can be stated that new definitions are free from the curvatures of utility curves. Strict concavity, strict convexity, and linearity imply the risk-averse, risk-loving, and risk-neutral investors, respectively, in standard theory. By contrast, the certain utility curve of a risk-averse investor, for instance, may be strictly concave or strictly convex, or linear for the new definitions. These properties of new definitions extend the standard utility theory. Hence, the new definitions make scientists' duties much easier in the real world and provide more flexible framework when modelling risk behaviors.

Alternative shapes of utility curves for new definitions may be tested empirically in financial models by the future works of researchers. These future works contribute to existing literature in important ways and may provide new areas for study.

## 6. Statements and Declarations

**Competing Interests:** The author has no relevant financial or non-financial interests to disclose.

**Data Availability Statement:** I do not analyze or generate any datasets, because my work proceeds within a theoretical and mathematical approach.

**Funding**: No funds, grants, or other support was received.